\documentclass[cameraready]{Interspeech}

\title{PASQA: Pitch-Accent-Focused Speech Quality Assessment Model \\ Trained on Synthetic Speech with Accent Errors}

\author[affiliation={1}]{Masaya}{Kawamura}
\author[affiliation={1}]{Yuma}{Shirahata}
\author[affiliation={1}]{Kentaro}{Mitsui}
\author[affiliation={1}]{Reo}{Shimizu}

\address{
    $^1$ LY Corporation, Japan
}

\email{kawamura.masaya, yuma.shirahata, kemitsui, reshimiz@lycorp.co.jp}


\usepackage{comment}
\usepackage{graphicx}
\usepackage{multirow}
\usepackage{amssymb}
\newcommand{\kedit}[1]{\textcolor{black}{#1}} 
\newcommand{\kkedit}[1]{\textcolor{black}{#1}} 
\newcommand{\kkkedit}[1]{\textcolor{black}{#1}} 
\newcommand{\sedit}[1]{\textcolor{black}{#1}}
\newcommand{\kkkkedit}[1]{\textcolor{black}{#1}} 
\newcommand{\medit}[1]{\textcolor{black}{#1}} 

\begin{document}

\fontsize{9.0}{10.6}\selectfont

\maketitle

\begin{abstract}
Existing mean opinion score (MOS) prediction models typically predict utterance-level naturalness \kkedit{MOS} and can be insensitive to localized pitch-accent errors. We propose Pitch-Accent-\kkkedit{f}ocused Speech Quality Assessment (PASQA)\kkedit{, which explicitly targets pitch-accent correctness.}
To train our model, we construct a controlled \kkedit{Japanese} accent-error \kkedit{dataset} by \kkkedit{changing accent patterns} using an accent-controllable \kkedit{text-to-speech} \medit{system}, and \kkkedit{compute a pseudo accent-quality score from the accent-error rate.}
\kkkkedit{
PASQA builds on self-supervised representations and employs mora-conditioned fusion, ranking loss, an auxiliary accent-error localization task, and speaker-invariant training.
}
Experiments show that conventional models fail to preserve the \kkkkedit{ordering by accent-error severity}, whereas PASQA achieves high ordering accuracy on both seen and unseen speakers.
\kkkedit{\kkkkedit{Further, }PASQA shows stronger agreement with human accent-correctness judgments.}
The \kkedit{code} is available at \url{https://github.com/lycorp-jp/PASQA}.
\end{abstract}

\section{Introduction}
\kkedit{Recent deep neural network (DNN)-based text-to-speech (TTS) systems can generate highly natural speech~\cite{NaturalSpeech3,chen2024vall,NEURIPS2023_2d8911db}.}
The quality of synthesized speech has conventionally been assessed using subjective listening tests, particularly Mean Opinion Score (MOS) evaluations by human raters\kkedit{, which provide accurate assessments}. However, such evaluations are costly and time-consuming\kkedit{~\cite{Erica_Cooper}}.
The MOS prediction \kkedit{models} \kkedit{have} therefore become increasingly used for rapid evaluation.
In recent years, with the advancement of \kkkkedit{DNNs} and the increase in available data, models that predict human preferences have been actively studied~\cite{45744,saeki22c_interspeech,dnsmos}.

These approaches typically estimate an utterance-level score intended to reflect overall naturalness.
However, speech naturalness is influenced not only by global signal quality but also by language-specific prosodic cues that carry lexical or grammatical information. 
\kkedit{For example,} \kkedit{i}n Japanese, pitch accent serves as such a perceptual cue~\cite{Ariga2025,Cutler1999}. 
Even slight shifts in the position of the accent nucleus can alter lexical meaning, making it crucial for intelligibility and naturalness \kkedit{(e.g., the Japanese word “hashi,” which means “chopsticks” or “bridge” depending on accent placement)}.
\kkedit{In contrast}, conventional utterance-level naturalness scores can be insensitive to such localized pitch-accent errors. 
This tendency is also observed in our experimental results (see Section~\ref{sec:results}). 
\kedit{\kkkkedit{Some} \kkedit{studies} have also explored fine-grained, frame-level quality prediction for synthetic speech to improve explainability and localization of degradations~\cite{kuhlmann25_interspeech}\kkedit{, but these approaches} have not yet focused on the correctness of pitch accent.} These findings motivate the need for an assessment \kedit{model} that explicitly targets pitch-accent correctness.

If a TTS system explicitly includes an accent prediction module, a straightforward approach would be to evaluate pitch-accent control by measuring the \kedit{module's accuracy}~\kkkkedit{\cite{park22b_interspeech,shirahata24_interspeech}}.  
\kedit{However, in many modern TTS architectures~\cite{du2024cosyvoice,chen-etal-2024-f5tts}, accent-related representations are not explicitly exposed and may be treated as black boxes.
In such cases, internal accent labels or intermediate prosodic predictions are unavailable at evaluation time.}
\kedit{Therefore, to ensure applicability across diverse TTS systems, accent-focused assessment should operate directly on the speech signal.}

To address \kedit{these problems}, we propose Pitch-Accent-focused Speech Quality Assessment (PASQA), \kkkkedit{a speech quality assessment model focused on }\kedit{pitch-accent correctness}. To enable \kkkkedit{learning and evaluation of accent correctness}, we first develop a corpus that covers representative pitch-accent error patterns.
\kkkkedit{Since real-world datasets rarely provide accent-error labels, we construct a Japanese dataset using controllable TTS to generate synthetic speech samples with controlled accent errors.}
\kkkkedit{
Our model builds on self-supervised representations, and to further enhance it, we adopt four strategies: additional mora-sequence inputs, ranking-based learning, an auxiliary task for accent-error localization, and speaker-invariant learning.
}

Experimental results show that conventional utterance-level MOS prediction models do not adequately reflect pitch-accent correctness.
In contrast, PASQA better preserves the \kkkkedit{ordering by accent-error severity} and shows stronger agreement with human \kedit{judgments of accent-correctness, }\kkkkedit{achieving a Spearman’s rank correlation coefficient (SRCC) of 0.828 and a Kendall’s $\tau$ (KTAU) of 0.614, both higher than those of conventional MOS prediction models.}
\kkkkedit{
These results indicate that explicitly modeling accent errors improves Japanese pitch-accent quality assessment.
}
\kkkkedit{
Moreover, PASQA demonstrates robust performance on an out-of-domain (OOD) TTS model.
}

\section{Proposed method}

\subsection{Accent-error \kedit{dataset}}\label{sec:accent-error-dataset}
\begin{figure}[t]
    \centering
    \includegraphics[width=0.45\textwidth]{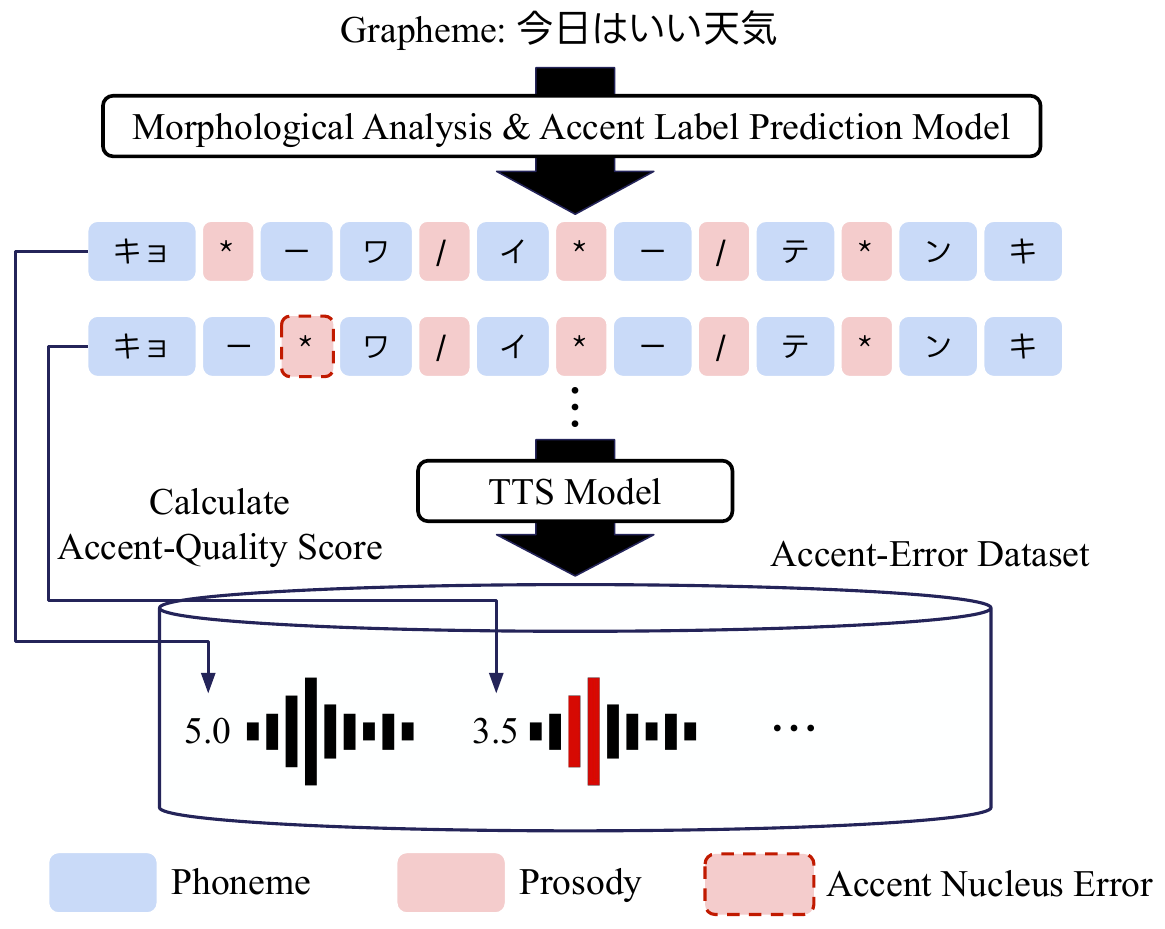}
    \vspace{-2mm}
    \caption{\kedit{Accent-error} \kkedit{dataset construction pipeline}.}
    \label{fig:pipeline}
\vspace{-6mm}
\end{figure}
To develop a model that reflects accent errors in its predicted accent-quality scores, we constructed a Japanese speech dataset that includes accent errors.
\kedit{Figure~\ref{fig:pipeline} shows} the overall pipeline for constructing the accent error \kedit{dataset}. 
In the figure, “/” denotes accent phrase boundaries\kkkedit{, which segment an utterance into prosodic units,} and “*” indicates the accent nucleus\kkkedit{, the mora where the pitch falls within an accent phrase.}
We use a TTS model with a DNN-based prosodic label prediction model~\cite{park22b_interspeech}.
It enables Japanese speech synthesis \kkedit{with} explicit control over accent patterns.

For each sentence, we derive prosodic annotations using the prosodic label prediction model.
\kkkedit{The annotations consist of three components: the mora sequence, accent phrase boundaries, and the accent nucleus.}

Accent errors are created by modifying the nucleus position in a subset of phrases.
Given a target error rate $r$, we uniformly sample $\max(1,\lfloor rP \rfloor)$ phrases from the $P$ accent phrases and alter their nucleus.
\kkkedit{
For a phrase of length $L$, valid accent types are 
$\{0, 1, \ldots, L-1\}$,
where 0 denotes the flat (0-type) accent and $k \in \{1, \ldots, L-1\}$ indicates a nucleus on the $k$-th mora. 
In Japanese pitch accent, a $\kkkkedit{k}$-type accent means that the pitch drops after the $\kkkkedit{k}$-th mora, whereas the 0-type (flat) accent has no pitch drop within the phrase.
}
\kedit{We uniformly resample the nucleus from valid positions, excluding the original, yielding unbiased accent-type conversions.}
The actual error rate is computed as the ratio of mora in modified phrases to the total number of mora in the utterance.
\kedit{Each sample is assigned a pseudo accent-quality score by applying a monotonic mapping to the actual error rate.}
\kkkkedit{
Specifically, we compute an utterance-level accent-quality score as
$S_{aq} = 5.0 - 4.0 \times \frac{N_{\mathrm{corr}}}{N}$,
where $N$ denotes the total number of mora in the utterance and
$N_{\mathrm{corr}}$ the number of mora belonging to corrupted accent phrases.
}

\subsection{PASQA}

\subsubsection{Model architecture}

\begin{figure}[t]
    \centering
    \includegraphics[width=0.43\textwidth]{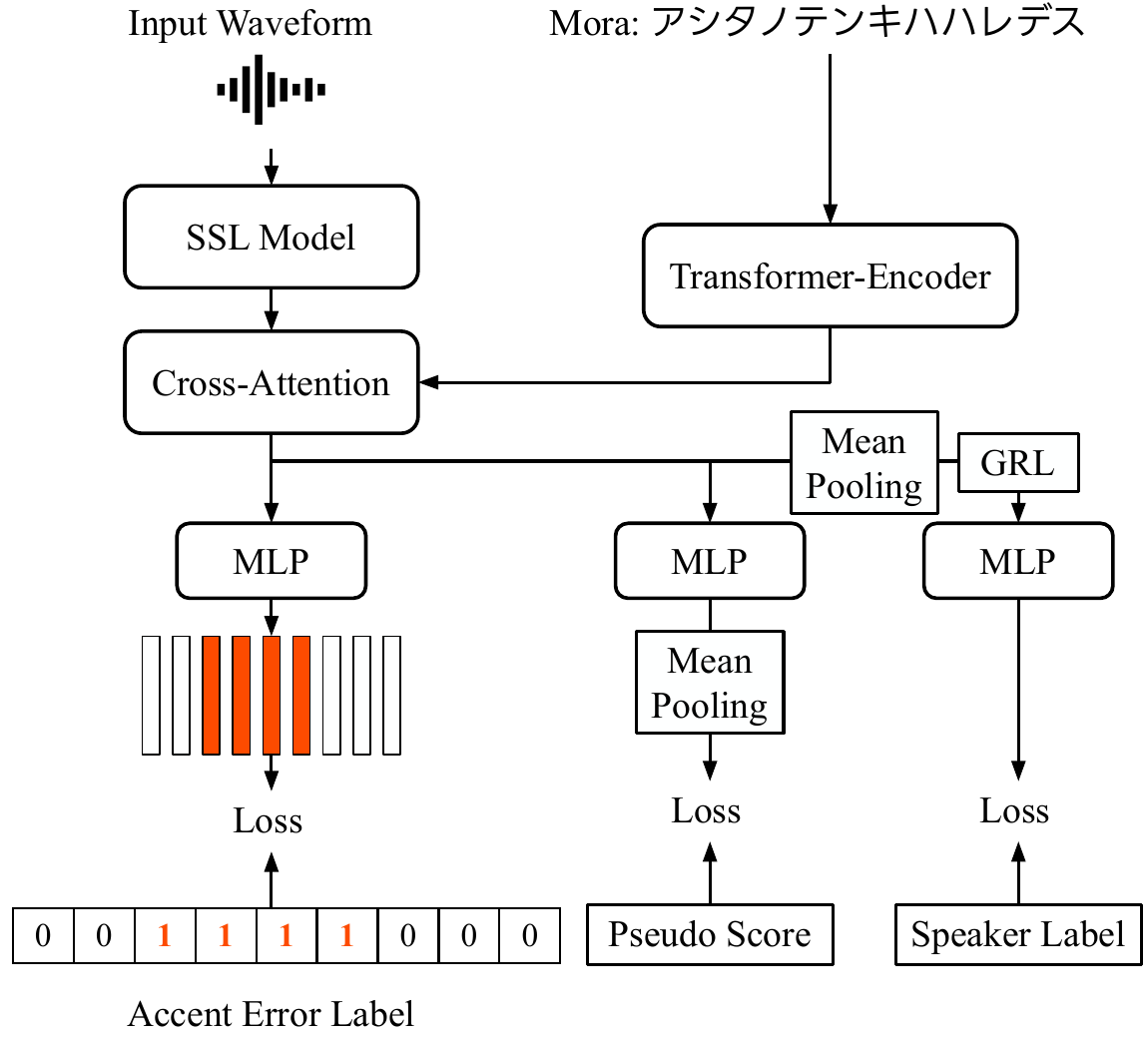}
    \vspace{-2mm}
    \caption{Overview of the proposed PASQA model.}
    \label{fig:accent-mos-overview}
    \vspace{-7mm}
\end{figure}

\kkedit{Figure~\ref{fig:accent-mos-overview} shows the overview of the proposed PASQA.}
\kedit{We adopt the SSL-MOS~\cite{sslmos} architecture as the backbone of our proposed model. 
SSL-MOS is a non-intrusive speech quality prediction framework that extracts self-supervised acoustic representations from waveforms and estimates an utterance-level score using a projection head with masked mean pooling.}
\kkkkedit{In PASQA, }the input is a waveform, and wav2vec 2.0~\cite{NEURIPS2020_92d1e1eb} produces frame-level acoustic \kedit{features}.
The downstream network predicts an accent-quality score.
To further improve prediction performance and robustness, we augment the base architecture with \kkkkedit{four} modifications: 1) mora-conditioned \kkkkedit{fusion} that incorporates the mora sequence as auxiliary linguistic information, 2) \kkkkedit{ranking loss}, 3) a frame-level accent-error detection head for auxiliary supervision, and 4) a \kedit{gradient reversal layer} (GRL)~\cite{grl} to encourage speaker-invariant representations. 
We describe each component in detail in the following subsections.

\subsubsection{Mora sequence as auxiliary linguistic information}
In the TTS evaluation setting, the input text is available.
\kkedit{
Since pitch accent in Japanese is defined at the mora level, we derive the mora sequence from the text and incorporate it as auxiliary linguistic information to explicitly model accent placement.}
\kkkedit{
Each utterance is represented by a mora sequence obtained from text analysis, which is then tokenized and embedded into a fixed-dimensional vector space.
}
We contextualize the sequence using a Transformer encoder~\cite{NIPS2017_3f5ee243}.
Mora information is fused with acoustic frames via cross-attention \kkkkedit{to generate mora-conditioned acoustic representations}.

\subsubsection{\kedit{Accent-quality head and ranking loss}}
\kkkedit{
The accent-quality \kkedit{head} is composed of a multilayer perceptron \kkedit{(MLP)}.
Range clipping maps outputs to the $[1,5]$ accent-quality \kkkkedit{score} interval using a tanh transform.
We train the accent-quality head using a pairwise logistic ranking loss,
inspired by the Bradley–Terry model~\cite{bradley1952rank}. This loss emphasizes ordinal relations:
\begin{equation}
P(i>j)=\sigma(\hat{y}_i-\hat{y}_j), \quad \mathcal{L}_{\mathrm{BT}}=-\sum_{i,j: y_i>y_j} \log P(i>j),
\end{equation}
where $\sigma$ denotes the sigmoid function, $\hat{y}_i$ is the predicted score for utterance $i$, and $y_i$ is the corresponding accent-quality score.
This objective learns relative ordering.
We first average frame-level scores over time to obtain an utterance-level score and then compute the loss over all unique pairs $(i,j)$ satisfying $y_i > y_j$ within a mini-batch, yielding $\frac{B(B-1)}{2}$ pairs for batch size $B$.
}

\subsubsection{Auxiliary frame error head}\label{sec:objective}
\kedit{Utterance-level scores do not explicitly reflect where pitch-accent errors occur along the temporal axis. 
To address this limitation, we introduce an auxiliary task that detects the temporal locations of pitch-accent errors to improve the estimation of the accent-quality score.}
\kkkedit{
Specifically, we add a frame-level auxiliary head that predicts accent-error labels from the outputs of the SSL models. 
Let $t = 1, \ldots, T$ index the encoder frames, where $T$ denotes the total number of frames. 
For each frame $t$, let $\kkkedit{l}_t \in \{0,1\}$ denote the corresponding binary label, 
where $\kkkedit{l}_t = 1$ indicates that frame $t$ belongs to an accent phrase whose nucleus has been modified, and $\kkkedit{l}_t = 0$ otherwise. 
The frame-level labels are obtained via alignment using the phoneme-level duration predictor of the TTS model.
We optimize a binary cross-entropy loss $\mathcal{L}_{\mathrm{frame}}$.
}

\subsubsection{\kedit{Speaker-invariant representation}}
\kedit{The model may exploit speaker-specific acoustic traits rather than accent-related cues.}
To reduce speaker-specific bias, we attach a speaker classifier to the utterance-level representation via a GRL~\cite{grl}.
\kedit{Specifically, we apply masked mean pooling to the SSL model's outputs to obtain an utterance embedding, which we then feed to the speaker classifier via GRL.}
The classifier is trained with cross-entropy loss, while the SSL model receives reversed gradients, encouraging speaker-invariant representations.
Inspired by~\cite{10901998}, we adapt scheduled GRL to our proposed model.
In detail, we scale the reversal with $\rho(p)=\frac{4}{1+\exp(-\gamma p)}-3$, \kedit{where \kkedit{$\gamma$ is a constant}, $p \in [0,1]$ denotes the normalized training progress, defined as the ratio of the current training step to the total number of training steps, and $\rho(p) \in [-1,1]$}.

\section{Experimental evaluation}
\subsection{Experimental setup}\label{sec:exp_setup}

\textbf{Dataset preparation.}
We generate synthetic Japanese speech with controlled pitch-accent errors using NANSY-TTS~\cite{choi2022nansy++}.
The TTS model was trained on an internal Japanese corpus consisting of 173,987 samples with manually-annotated phonemic and prosodic labels, totaling 207.96 hours. 
This corpus included 17 speakers.
\kedit{We apply this TTS model to 91{,}157 sentences to construct a dataset containing controlled pitch-accent errors.}
\kedit{Prosodic annotations are obtained using a MeCab-based morphological analysis model~\cite{kudo-etal-2004-applying} for text normalization, followed by a DNN-based prosodic label prediction model~\cite{park22b_interspeech}. 
\kkedit{The prosodic label prediction model is trained on 80{,}061 manually annotated prosodic labels.}
Further details can be found in~\cite{park22b_interspeech}. 
These annotations are used to manipulate accent nucleus for controlled data construction. 
This process also provides mora sequences, which are used as auxiliary linguistic inputs in our model.
}

For each sentence, to evaluate how sensitively the model responds to the proportion of accent errors, 
\kkkedit{
we explicitly manipulate the accent nucleus and construct three severity conditions corresponding to different accent-error rates $r$. 
Specifically, we generate speech with $r = 0$ for the error-free condition, with \sedit{$r \in [0.1, 0.2]$} 
for the low-severity condition, and with \sedit{$r \in [0.8,0.9]$} 
for the high-severity condition\kkkkedit{.} 
}
Therefore, three distinct speech samples are generated for each utterance.
We generated these samples for 13 speakers.
\kedit{All generated samples were split into 80\% for training and 20\% for development. 
As a result, the training set consists of 2,130,85\kkkkedit{8} speech samples, with a total duration of 2,898.79 hours, and the remaining samples were used for validation.}
\kedit{For the test set, we prepared 1,170 speech samples in total from 13 seen speakers used in training and 2,400 speech samples in total from 4 unseen speakers. 
}

\noindent\textbf{Model details.}
The backbone of our model follows the SSL-MOS~\cite{sslmos} architecture with wav2vec 2.0~\cite{NEURIPS2020_92d1e1eb} frame-level features and a scalar prediction head. The utterance-level score is predicted by a \kkedit{two-layer} \kkedit{MLP} with a hidden size of 64 and range clipping. For mora-conditioned variants, mora tokens are embedded in 256 dimensions, contextualized by a 1-layer Transformer encoder (4 heads, feed forward network dimension 512, dropout 0.1) with rotary positional encoding\kkkkedit{~\cite{SU2024127063}}, and fused with acoustic features via cross-attention (attention dimension 256, 4 heads, dropout 0.1). We further add an auxiliary frame-level error head (hidden size 64) and a speaker-adversarial branch with a GRL (speaker classifier hidden size 128, dropout 0.1).

\noindent\textbf{Model training.}
All models are trained on 16~kHz waveforms.
The model is optimized using stochastic gradient descent with a learning rate of $1\times10^{-3}$ and momentum of 0.9, and a batch size of 16.
We apply gradient clipping with norm 1.0 and train for up to 100{,}000 steps.
The final loss is as follows\sedit{:}
\begin{equation}
\mathcal{L}=\kkkkedit{\lambda_{\mathrm{BT}}\,}\mathcal{L}_{\mathrm{BT}}+\kedit{\lambda_{\mathrm{L1}}\,\mathcal{L}_{\mathrm{L1}}}+\lambda_{\mathrm{frame}}\,\mathcal{L}_{\mathrm{frame}}+\lambda_{\mathrm{spk}}\,\mathcal{L}_{\mathrm{spk}}.
\end{equation}
\kkkkedit{Following the original SSL-MOS formulation,}
\kkedit{we also include an L1 loss $\mathcal{L}_{\mathrm{L1}}$ between the predicted utterance-level score and the target audio quality score.}
\kkkkedit{The loss weights $\lambda_{\mathrm{BT}}$, $\lambda_{\mathrm{L1}}$, $\lambda_{\mathrm{frame}}$, and $\lambda_{\mathrm{spk}}$ were set to 1.5, 0.5, 0.2, and 0.1, respectively.
For GRL models, we use a GRL schedule with $\gamma=10$.
}

\noindent\textbf{Comparison methods.}
We compare against widely used non-intrusive quality predictors, including DNSMOS P.835~\cite{9746108}, DNSMOS P.808~\cite{naderi20_interspeech}\kedit{,} and NISQA~\cite{mittag21_interspeech}.
We also use UTMOS~\cite{saeki22c_interspeech}, UTMOSv2~\cite{10832315} and \kedit{SHEET SSL-MOS}~\cite{sheet,huang2024mos}.
We obtain these models' scores using the VERSA toolkit~\cite{shi2025versa,shi2024versaversatileevaluationtoolkit}.
\kedit{As models trained on the accent-error dataset, we train two baseline models and the proposed PASQA.}
To evaluate the effectiveness of SSL-based representations, \kedit{we compare against a model trained on features extracted using WORLD~\cite{morise2016world}, \kkkedit{referred to as ACC-WORLD-MOS}.}
WORLD features are extracted at a 10~ms frame period from 16~kHz audio. We estimate $f_0$ within 50--500~Hz and use log-$f_0$, a voiced/unvoiced flag, 24-dimensional mel-cepstral coefficients, and a 1-dimensional aperiodicity feature, concatenated into a 27-dimensional frame-level representation.
We also include an SSL-MOS model trained solely with an L1 loss as a baseline, \kkkedit{referred to as ACC-SSL-MOS}.

\noindent\textbf{Evaluation metrics.}
\kedit{To verify that the model can differentiate between minimally and severely corrupted accent patterns within the same utterance, we evaluate its ability to preserve severity ordering. }
Each sentence has three controlled severity conditions. We assess whether predicted scores preserve the \kkkkedit{expected ordering, i.e., error-free $>$ low-severity $>$ high-severity}. 
Order accuracy is defined as the fraction of triplets that satisfy this strict ordering. 
Triplets are formed from the three variants synthesized from the same text and speaker, and ties are treated as violations. 
In addition, to assess how well the predicted utterance-level scores align with the accent-quality \kedit{score} derived, we \kedit{measure} Pearson's linear correlation coefficient (LCC), \kkkkedit{SRCC}, and \kkkkedit{KTAU}.

\begin{table*}[t]
\centering
\caption{Results on seen and unseen speakers in the accent-error evaluation set.}
\vspace{-4mm}
\label{tab:accent_results}
\setlength{\tabcolsep}{4pt}
\scalebox{0.90}{
\begin{tabular}{c l cccc|cccc}
\toprule
\multirow{2}{*}{Accent-error dataset} 
& \multirow{2}{*}{Model}
& \multicolumn{4}{c|}{Seen speaker} 
& \multicolumn{4}{c}{Unseen speaker} \\
\cline{3-10}
& & Order Acc. & LCC & SRCC & KTAU
& Order Acc. & LCC & SRCC & KTAU \\
\hline

\multirow{6}{*}{\kkkedit{No}} 
& DNSMOS P.835
& 0.200 & -0.001 & -0.001 & -0.001
& 0.121 & -0.073 & -0.057 & -0.040 \\

& DNSMOS P.808
& 0.167 & -0.031 & -0.028 & -0.019
& 0.170 & -0.029 & -0.025 & -0.017 \\

& NISQA
& 0.156 & -0.010 & 0.011 & 0.007
& 0.195 & 0.004 & 0.006 & 0.004 \\

& SHEET SSL-MOS
& 0.139 & -0.017 & -0.049 & -0.035
& 0.174 & -0.004 & -0.029 & -0.020 \\

& UTMOS 
& 0.133 & -0.020 & -0.013 & -0.009
& 0.121 & -0.039 & -0.041 & -0.029 \\

& UTMOSv2 
& 0.162 & -0.009 & -0.009 & -0.006
& 0.134 & -0.054 & -0.047 & -0.033 \\

\hline

\multirow{7}{*}{\kkkedit{Yes}}
& \kkkedit{ACC-WORLD-MOS}
& 0.346 & 0.020 & 0.036 & 0.025
& 0.339 & 0.040 & 0.040 & 0.028 \\

& \kkkedit{ACC-}SSL-MOS 
& 0.710 & 0.811 & 0.666 & 0.472 
& 0.738 & 0.818 & 0.724 & 0.530 \\

& PASQA
& \textbf{0.754} & 0.829 & \textbf{0.711} & \textbf{0.524}
& 0.785 & \textbf{0.879} & \textbf{0.751} & \textbf{0.559} \\

& PASQA w/o bradley-terry loss 
& 0.723 & \textbf{0.837} & 0.678 & 0.478
& \textbf{0.787} & 0.863 & 0.742 & 0.547 \\

& PASQA w/o frame error head 
& 0.721 & 0.836 & 0.658 & 0.463
& 0.747 & 0.838 & 0.720 & 0.527 \\

& PASQA w/o GRL 
& 0.662 & 0.779 & 0.625 & 0.433
& 0.745 & 0.818 & 0.712 & 0.520 \\

& PASQA w/o mora-conditioned fusion 
& 0.695 & 0.761 & 0.628 & 0.441
& 0.735 & 0.833 & 0.721 & 0.526 \\

\bottomrule
\vspace{-8mm}
\end{tabular}
}
\end{table*}

\noindent\textbf{Subjective evaluation.}
We conducted a listening test with 15 native Japanese speakers.
\kedit{The listening test included 120 speech samples synthesized from four speakers (two male and two female).}
\kkedit{These speakers were included in the training dataset.
The text used for synthesis did not overlap with those in the training dataset.
}
Participants were instructed to rate each sample on a five-point scale based on whether its pitch accent sounded natural in the Tokyo dialect. We compute order accuracy on the aggregated ratings. We also measure \sedit{mean squared error (MSE)}, LCC, SRCC, and KTAU between model predictions and the aggregated human ratings.

\noindent\textbf{\kkkedit{Out-of-domain evaluation.}} 
\kkedit{To verify whether PASQA performs robustly on speech synthesized by an \kkkkedit{OOD} TTS model, we use GPT-4o-mini-TTS~\cite{gpt}.
We prepared 50 texts that do not overlap with the training dataset and synthesized speech from them.
We synthesized \sedit{speech using} two input patterns: grapheme and mora sequence.
In a preliminary listening test, we confirmed that speech synthesized from grapheme input tended to exhibit better Japanese pitch-accent quality than that synthesized from mora input.
Based on this observation, we assigned the grapheme-input sample as having higher accent quality and treated it as the positive label.
We then computed pairwise accuracy based on both the model predictions and the judgments of 10 \kkkedit{native Japanese speakers} regarding accent naturalness.}

\subsection{Experimental results}\label{sec:results}

\begin{table}[t]
\centering
\caption{Subjective evaluation results.}
\vspace{-4mm}
\setlength{\tabcolsep}{6pt}
\scalebox{0.90}{
\begin{tabular}{l@{\hspace{5pt}}c@{\hspace{4pt}}c@{\hspace{5pt}}c@{\hspace{5pt}}c@{\hspace{5pt}}c}
\toprule
Model & Order Acc. & MSE & LCC & SRCC & KTAU \\
\hline
Human ratings & 0.925 & - & - & - & - \\\hline
DNSMOS P.835& 0.175 & \textbf{0.811} & -0.074 & -0.074 & -0.053 \\
DNSMOS P.808& 0.150 & 1.272 & -0.043 & -0.034 & -0.027\\
NISQA & 0.175 & 1.593 & -0.012 & -0.028 & -0.022 \\
SHEET SSL-MOS & 0.125 & 3.059 & -0.038 & -0.058 & -0.042 \\
UTMOS & 0.125 & 0.943 & 0.007 & -0.012 & -0.008 \\
UTMOSv2 & 0.075 & 1.151 & -0.158 & -0.171 & -0.117  \\\hline
\kkkedit{ACC-WORLD-MOS} & 0.350 & 1.945&  0.111& 0.142  &0.095\\
\kkkedit{ACC-}SSL-MOS & \textbf{0.900} & 1.272 & 0.739 & 0.764 & 0.541 \\
PASQA & 0.850 & 1.293 & \textbf{0.814} & \textbf{0.828} & \textbf{0.614} \\
\bottomrule
\end{tabular}
}
\label{tab:subjective_results}
\vspace{-4mm}
\end{table}

Table~\ref{tab:accent_results} shows objective evaluation results on the controlled accent-error dataset, where each text prompt is synthesized into three severity conditions (error-free, low\kkkkedit{-severity}, high\kkkkedit{-severity}) and we measure whether predicted utterance-level scores preserve the expected ordering. \kkkkedit{Publicly available models that are not trained on the accent-error dataset} yield near-chance ordering and \kkedit{correlations close to zero, often negative, suggesting} that their utterance-level naturalness scores are not aligned with pitch-accent correctness under localized nucleus errors.

In contrast, models trained on the accent-error \kkedit{dataset} substantially improve both ordering and proxy correlation. 
\kkedit{This result indicates that the \kkkkedit{constructed} accent-error dataset is effective for training pitch-accent quality assessment models.}
\kkkedit{ACC-WORLD-MOS} improves \kkkedit{order accuracy on both seen and unseen speakers} but remains weak in correlation. \kkkedit{ACC-SSL-MOS substantially outperforms ACC-WORLD-MOS across all evaluation metrics.} 
\kkkkedit{
This suggests that data-driven self-supervised representations capture richer prosodic cues related to accent-error severity than acoustic features such as WORLD parameters.
}
\kkkedit{PASQA further outperformed both ACC-WORLD-MOS and ACC-SSL-MOS across all metrics for both seen and unseen speakers. These results suggest that PASQA, including the auxiliary linguistic information, ranking loss, frame error head, and GRL, effectively enhanced the performance of accent-quality assessment.}

Ablation results indicate complementary contributions from each component. Removing the frame-level error head or mora-conditioned fusion degrades ordering and correlation, consistent with the role of localized supervision and linguistic conditioning for detecting mild errors. Removing GRL most strongly affects the seen-speaker condition, suggesting that speaker-adversarial training helps mitigate speaker-specific bias in the controlled corpus.

\kkedit{Table~\ref{tab:subjective_results} shows that human ratings show high consistency in preserving the expected ordering (error-free $>$ low $>$ high), achieving an order accuracy of 0.925.}
\kkedit{
\kkkedit{ACC-}SSL-MOS achieves the highest order accuracy, while PASQA achieves the strongest agreement with human ratings in terms of LCC, SRCC, and KTAU.
These results suggest that the architectural enhancements in PASQA, including the ranking loss, auxiliary frame-level error head, GRL, and auxiliary linguistic information, contribute to improved robustness and higher correlation with human ratings.
}
\kkedit{The conventional MOS prediction models achieve lower MSE than PASQA. 
This is likely because PASQA is trained on pseudo accent-quality scores, which may lead to a mismatch between the dynamic range of predicted scores and the scale of human ratings. 
However, the primary objective of this study is not absolute score calibration but \kkkkedit{accurate severity ordering} and sensitivity to localized accent errors. 
}

\kkedit{
Table~\ref{tab:pairwise_binom} shows pairwise accent-quality discrimination between speech synthesized from grapheme input and mora-sequence input using GPT-4o-mini-TTS.
PASQA achieves the highest pairwise accuracy and significantly exceeds chance level, while conventional MOS predictors fail to reach statistical significance. 
This result indicates that PASQA performs robustly on \kkkedit{OOD} TTS systems \kkkkedit{and} remain\kkkkedit{s} sensitive to accent-quality differences.
}

\begin{table}[t]
\centering
\caption{
\kkedit{Pairwise accuracy on GPT-4o-mini-TTS outputs.
}
\kkkedit{$p$-values are computed using a one-sided exact binomial test against chance level (0.5).}
}
\vspace{-4mm}
\label{tab:pairwise_binom}
\scalebox{0.90}{
\begin{tabular}{lcc}
\toprule
Model & Pairwise Acc. & $p$-value \\
\midrule
Human ratings &\kkkedit{0.984} &\kkkedit{$<0.001$} \\\hline
DNSMOS P.835      & 0.380 & 0.968  \\
DNSMOS P.808         & 0.320 & 0.997  \\
NISQA  & 0.620 & 0.060 \\
SHEET SSL-MOS       & 0.520 & 0.444  \\
UTMOS             & 0.260 & 0.999 \\
UTMOSv2           & 0.580 & 0.161  \\\hline
\kkkedit{ACC-WORLD-MOS}  & 0.420 & 0.899 \\
\kkkedit{ACC-}SSL-MOS &0.720 & 0.001 \\
PASQA    & \textbf{0.780} & \kkkedit{$<0.001$} \\
\bottomrule

\end{tabular}
}
\vspace{-5mm}
\end{table}

\section{Conclusion}
\kkkedit{
We proposed PASQA, a pitch-accent-focused speech quality assessment model.
Using a controllable TTS system, we constructed a scalable accent-error dataset without manual annotation.
Built on SSL-based acoustic representations, PASQA improves accent-quality assessment and outperforms conventional MOS models.
In listening tests, it also shows stronger agreement with human accent-correctness judgments.
Future work will focus on improving robustness to OOD scenarios and extending the framework to multilingual settings.
}

\section{Generative AI Use Disclosure}
\kkkedit{In accordance with ISCA policy, generative AI tools were used solely for English language editing and polishing of the manuscript.
All (co-)authors have reviewed the final version and are fully responsible and accountable for the scientific content, experimental design, results, and conclusions.
}
\bibliographystyle{IEEEtran}
\bibliography{mybib}

@inproceedings{choi2022nansy++,
  title={{NANSY++}: Unified Voice Synthesis with Neural Analysis and Synthesis},
  author={Choi, Hyeong-Seok and Yang, Jinhyeok and Lee, Juheon and Kim, Hyeongju},
  booktitle={Proc. ICLR},
  year={2023},
}

@inproceedings{sheet,
  title     = {{SHEET}: {A} Multi-purpose Open-source Speech Human Evaluation Estimation Toolkit},
  author    = {Wen-Chin Huang and Erica Cooper and Tomoki Toda},
  year      = {2025},
  booktitle = {{Proc. Interspeech}},
  pages     = {2355--2359},
}

@inproceedings{shi2025versa,
title={{VERSA}: A Versatile Evaluation Toolkit for Speech, Audio, and Music},
author={Jiatong Shi and Shim, Hye-jin and Jinchuan Tian and Siddhant Arora and Haibin Wu and Darius Petermann and Jia Qi Yip and You Zhang and Yuxun Tang and Wangyou Zhang and Dareen Safar Alharthi and Yichen Huang and Koichi Saito and Jionghao Han and Yiwen Zhao and Chris Donahue and Shinji Watanabe},
booktitle={Proc. NAACL-HLT (System Demonstrations)},
year={2025},
pages = "191--209"
}

@inproceedings{shi2024versaversatileevaluationtoolkit,
  author={Shi, Jiatong and Tian, Jinchuan and Wu, Yihan and Jung, Jee-Weon and Yip, Jia Qi and Masuyama, Yoshiki and Chen, William and Wu, Yuning and Tang, Yuxun and Baali, Massa and Alharthi, Dareen and Zhang, Dong and Deng, Ruifan and Srivastava, Tejes and Wu, Haibin and Liu, Alexander and Raj, Bhiksha and Jin, Qin and Song, Ruihua and Watanabe, Shinji},
  booktitle={Proc. SLT}, 
  title={{ESPnet-Codec}: Comprehensive Training and Evaluation of Neural Codecs For Audio, Music, and Speech}, 
  year={2024},
  pages={562-569},
  doi={10.1109/SLT61566.2024.10832289}
}

@inproceedings{saeki22c_interspeech,
  title     = {{UTMOS}: {UTokyo}-SaruLab System for {VoiceMOS} Challenge 2022},
  author    = {Takaaki Saeki and Detai Xin and Wataru Nakata and Tomoki Koriyama and Shinnosuke Takamichi and Hiroshi Saruwatari},
  year      = {2022},
  booktitle = {Proc. Interspeech},
  pages     = {4521--4525},
  doi       = {10.21437/Interspeech.2022-439},
  issn      = {2958-1796},
}

@article{morise2016world,
  author  = {Masanori Morise and Fumiya Yokomori and Kenji Ozawa},
  title   = {{WORLD}: A Vocoder-Based High-Quality Speech Synthesis System for Real-Time Applications},
  journal = {IEICE Trans. Inf. Syst.},
  volume  = {E99-D},
  number  = {7},
  pages   = {1877--1884},
  year    = {2016},
}

@inproceedings{dnsmos,
  author={Reddy, Chandan K A and Gopal, Vishak and Cutler, Ross},
  booktitle={Proc. ICASSP}, 
  title={{DNSMOS}: {A} Non-Intrusive Perceptual Objective Speech Quality Metric to Evaluate Noise Suppressors}, 
  year={2021},
  volume={},
  number={},
  pages={6493-6497},}

@inproceedings{mittag21_interspeech,
  title     = {{NISQA}: A Deep {CNN}-Self-Attention Model for Multidimensional Speech Quality Prediction with Crowdsourced Datasets},
  author    = {Gabriel Mittag and Babak Naderi and Assmaa Chehadi and Sebastian Möller},
  year      = {2021},
  booktitle = {Proc. Interspeech},
  pages     = {2127--2131},
  doi       = {10.21437/Interspeech.2021-299},
  issn      = {2958-1796},
}

@inproceedings{10832315,
  author={Baba, Kaito and Nakata, Wataru and Saito, Yuki and Saruwatari, Hiroshi},
  booktitle={Proc. SLT}, 
  title={The {T05} System for the {VoiceMOS challenge} 2024: Transfer Learning from Deep Image Classifier to Naturalness {MOS} Prediction of High-Quality Synthetic Speech}, 
  year={2024},
  volume={},
  number={},
  pages={818-824},
  doi={10.1109/SLT61566.2024.10832315}}

@inproceedings{kuhlmann25_interspeech,
  title     = {Towards Frame-level Quality Predictions of Synthetic Speech},
  author    = {Michael Kuhlmann and Fritz Seebauer and Petra Wagner and Reinhold Haeb-Umbach},
  year      = {2025},
  booktitle = {Proc. Interspeech},
  pages     = {2300--2304},
  issn      = {2958-1796},
}

@article{grl,
  author={Abdelwahab, Mohammed and Busso, Carlos},
  journal={IEEE/ACM Trans. on Audio, Speech, and Lang. Process.}, 
  title={Domain Adversarial for Acoustic Emotion Recognition}, 
  year={2018},
  volume={26},
  number={12},
  pages={2423-2435},
  doi={10.1109/TASLP.2018.2867099}}

@inproceedings{sslmos,
  author={Cooper, Erica and Huang, Wen-Chin and Toda, Tomoki and Yamagishi, Junichi},
  booktitle={Proc. ICASSP}, 
  title={Generalization Ability of MOS Prediction Networks}, 
  year={2022},
  volume={},
  number={},
  pages={8442-8446},
  doi={10.1109/ICASSP43922.2022.9746395}}

@inproceedings{park22b_interspeech,
  title     = {A Unified Accent Estimation Method Based on Multi-Task Learning for {Japanese} Text-to-Speech},
  author    = {Byeongseon Park and Ryuichi Yamamoto and Kentaro Tachibana},
  year      = {2022},
  booktitle = {Proc. Interspeech},
  pages     = {1931--1935},
  doi       = {10.21437/Interspeech.2022-334},
  issn      = {2958-1796},
}

@inproceedings{NEURIPS2020_92d1e1eb,
 author = {Baevski, Alexei and Zhou, Yuhao and Mohamed, Abdelrahman and Auli, Michael},
 booktitle = {Proc. NeurIPS},
 pages = {12449--12460},
 title = {wav2vec 2.0: A Framework for Self-Supervised Learning of Speech Representations},
 volume = {33},
 year = {2020}
}

@article{SU2024127063,
title = {{RoFormer}: {Enhanced} transformer with Rotary Position Embedding},
journal = {Neurocomputing},
volume = {568},
pages = {127063},
year = {2024},
issn = {0925-2312},
doi = {https://doi.org/10.1016/j.neucom.2023.127063},
author = {Jianlin Su and Murtadha Ahmed and Yu Lu and Shengfeng Pan and Wen Bo and Yunfeng Liu}
}

@inproceedings{NIPS2017_3f5ee243,
 author = {Vaswani, Ashish and Shazeer, Noam and Parmar, Niki and Uszkoreit, Jakob and Jones, Llion and Gomez, Aidan N and Kaiser, \L ukasz and Polosukhin, Illia},
 booktitle = {Proc. NeurIPS},
 pages = {},
 title = {Attention is All you Need},
 volume = {30},
 year = {2017}
}

@article{bradley1952rank,
 ISSN = {00063444, 14643510},
 author = {Ralph Allan Bradley and Milton E. Terry},
 journal = {Biometrika},
 number = {3/4},
 pages = {324--345},
 publisher = {[Oxford University Press, Biometrika Trust]},
 title = {Rank Analysis of Incomplete Block Designs: I. The Method of Paired Comparisons},
 volume = {39},
 year = {1952}
}

@inproceedings{kudo-etal-2004-applying,
    title = "Applying Conditional Random Fields to {J}apanese Morphological Analysis",
    author = "Kudo, Taku  and
      Yamamoto, Kaoru  and
      Matsumoto, Yuji",
    booktitle = "Proc. EMNLP",
    year = "2004",
    pages = "230--237"
}

@inproceedings{naderi20_interspeech,
  title     = {An Open Source Implementation of {ITU-T} Recommendation {P.808} with Validation},
  author    = {Babak Naderi and Ross Cutler},
  year      = {2020},
  booktitle = {Proc. Interspeech},
  pages     = {2862--2866},
  doi       = {10.21437/Interspeech.2020-2665},
  issn      = {2958-1796},
}

@article{huang2024mos,
  title={{MOS-Bench}: Benchmarking generalization abilities of subjective speech quality assessment models},
  author={Huang, Wen-Chin and Cooper, Erica and Toda, Tomoki},
  journal={arXiv preprint arXiv:2411.03715},
  year={2024}
}

@ARTICLE{10901998,
  author={Qu, Leyuan and Weber, Cornelius and Wang, Wei and Jin, Jia and Gao, Yingming and Li, Taihao and Wermter, Stefan},
  journal={IEEE Trans. Neural Netw. Learn. Syst.}, 
  title={Disentanglement of Prosody Representations via Diffusion Models and Scheduled Gradient Reversal}, 
  year={2025},
  volume={36},
  number={8},
  pages={15043-15054},
  doi={10.1109/TNNLS.2025.3534822}}

@INPROCEEDINGS{9746108,
  author={Reddy, Chandan K A and Gopal, Vishak and Cutler, Ross},
  booktitle={Proc. ICASSP}, 
  title={{DNSMOS} {P.835}: A Non-Intrusive Perceptual Objective Speech Quality Metric to Evaluate Noise Suppressors}, 
  year={2022},
  volume={},
  number={},
  pages={886-890},
  doi={10.1109/ICASSP43922.2022.9746108}}

@article{Ariga2025,
  title = {Recognition of spoken words with mispronounced lexical prosody in {Japanese}},
  volume = {157},
  ISSN = {1520-8524},
  number = {6},
  journal = {J. Acoust. Soc. Am.},
  author = {Ariga,  Terumichi and Hirose,  Yuki},
  year = {2025},
  pages = {4102–4118}
}

@article{Cutler1999,
  title = {Pitch accent in spoken-word recognition in {Japanese}},
  volume = {105},
  ISSN = {1520-8524},
  number = {3},
  journal = {J. Acoust. Soc. Am.},
  author = {Cutler,  Anne and Otake,  Takashi},
  year = {1999},
  pages = {1877–1888}
}

@inproceedings{45744,title	= {{AutoMOS}: {Learning} a non-intrusive assessor of naturalness-of-speech},author	= {Brian Patton and Yannis Agiomyrgiannakis and Michael Terry and Kevin Wilson and Rif A. Saurous and D. Sculley},year	= {2016},booktitle	= {Proc. NeurIPS 2016 End-to-end Learning for Speech and Audio Processing Workshop}}

@article{du2024cosyvoice,
  title={{CosyVoice} 2: Scalable streaming speech synthesis with large language models},
  author={Du, Zhihao and Wang, Yuxuan and Chen, Qian and Shi, Xian and Lv, Xiang and Zhao, Tianyu and Gao, Zhifu and Yang, Yexin and Gao, Changfeng and Wang, Hui and others},
  journal={arXiv preprint arXiv:2412.10117},
  year={2024}
}

@inproceedings{chen-etal-2024-f5tts,
    title = "F5-{TTS}: A Fairytaler that Fakes Fluent and Faithful Speech with Flow Matching",
    author = "Chen, Yushen and
      Niu, Zhikang and
      Ma, Ziyang and
      Deng, Keqi and
      Wang, Chunhui and
      Zhao, Jian and
      Yu, Kai and
      Chen, Xie",
    booktitle = "Proc. ACL",
    year = "2025",
    pages = "6255--6271",
}

@article{Erica_Cooper,
  title={A review on subjective and objective evaluation of synthetic speech},
  author={Erica Cooper and Wen-Chin Huang and Yu Tsao and Hsin-Min Wang and Tomoki Toda and Junichi Yamagishi},
  journal={Acoustical Science and Technology},
  volume={45},
  number={4},
  pages={161-183},
  year={2024},
  doi={10.1250/ast.e24.12}
}

@inproceedings{NaturalSpeech3,
author = {Ju, Zeqian and Wang, Yuancheng and Shen, Kai and Tan, Xu and Xin, Detai and Yang, Dongchao and Liu, Yanqing and Leng, Yichong and Song, Kaitao and Tang, Siliang and Wu, Zhizheng and Qin, Tao and Li, Xiang-Yang and Ye, Wei and Zhang, Shikun and Bian, Jiang and He, Lei and Li, Jinyu and Zhao, Sheng},
title = {Natural{S}peech 3: {Z}ero-shot speech synthesis with factorized codec and diffusion models},
year = {2024},
booktitle = {Proc. ICML},
articleno = {909},
numpages = {19},
}

@article{chen2024vall,
  title={{VALL-E} 2: Neural codec language models are human parity zero-shot text to speech synthesizers},
  author={Chen, Sanyuan and Liu, Shujie and Zhou, Long and Liu, Yanqing and Tan, Xu and Li, Jinyu and Zhao, Sheng and Qian, Yao and Wei, Furu},
  journal={arXiv preprint arXiv:2406.05370},
  year={2024}
}

@inproceedings{NEURIPS2023_2d8911db,
 author = {Le, Matthew and Vyas, Apoorv and Shi, Bowen and Karrer, Brian and Sari, Leda and Moritz, Rashel and Williamson, Mary and Manohar, Vimal and Adi, Yossi and Mahadeokar, Jay and Hsu, Wei-Ning},
 booktitle = {Proc. NeurIPS},
 pages = {14005--14034},
 title = {Voicebox: {T}ext-Guided Multilingual Universal Speech Generation at Scale},
 volume = {36},
 year = {2023}
}

@article{gpt,
  title={Introducing next-generation audio models in the {API}},
  author={OpenAI},
  year={2025}
}

@inproceedings{shirahata24_interspeech,
  title     = {{Audio-conditioned phonemic and prosodic annotation for building text-to-speech models from unlabeled speech data}},
  author    = {Yuma Shirahata and Byeongseon Park and Ryuichi Yamamoto and Kentaro Tachibana},
  year      = {2024},
  booktitle = {Proc. Interspeech},
  pages     = {2795--2799},
  doi       = {10.21437/Interspeech.2024-342},
  issn      = {2958-1796},
}

\end{document}